\documentclass{ws-procs9x6-cpt22}

\usepackage[hidelinks]{hyperref}
\usepackage{mathtools}

\DeclarePairedDelimiterX\Braket[2]{\langle}{\rangle}{#1 \delimsize\vert #2}

\begin{document}
	

\newcommand\cf{\mbox{c.\,f.}\xspace}
\newcommand\eg{\mbox{e.\,g.}\xspace}
\newcommand\DX{\textrm{d}x}
\newcommand{\hx}{\hat{x}}
\newcommand{\hp}{\hat{p}}
\newcommand{\hX}{\hat{X}}
\newcommand{\hP}{\hat{P}}
\newcommand{\ie}{\mbox{\textit{i.\,e.}}\xspace}	
	
	\title{Curved momentum space equivalent to the linear and quadratic Generalized Uncertainty Principle}
	
	\author{F.\ Wagner,$^1$}
	
	\address{$^1$Institute of Physics, University of Szczecin\\ Wielkopolska 15, 70-451 Szczecin, Poland}

\begin{abstract}
		In this work, we deepen the correspondence between Generalized Uncertainty Principles (GUPs) and quantum dynamics on curved momentum space. In particular, we investigate the linear and quadratic GUP. Similarly to earlier work, the resulting curvature tensor in the dual theory is proportional to the coordinate non-commutativity of the original formulation.
\end{abstract}
	
\bodymatter

\section{Introduction}

In the absence of a fully convincing theory of quantum gravity (for a recent overview see \cite{Loll:2022ibq}) and with increasing experimental precision, the interest of a part of the community has shifted towards quantum spacetime phenomenology \cite{Addazi:2021xuf}. One of the historically most influential ideas in this arena is the minimum length paradigm \cite{Mead:1964zz}, be it incorporated in relativistic kinematics as in Deformed Special Relativity (DSR) \cite{Amelino-Camelia:2000stu} or at the quantum level by virtue of the Generalized Uncertainty Principle (GUP) \cite{Kempf:1994su,Maggiore:1993rv,Bosso:2022vlz,Bosso:2020aqm,Buoninfante:2019fwr}.

It has been known for some time that DSR may be understood as a theory of de Sitter-shaped momentum space \cite{Kowalski-Glikman:2003qjp}. Recently, it has become increasingly clear that an analogous correspondence exists between deformed Heisenberg algebras and curved momentum \cite{Wagner21a,Wagner22,Singh:2021iqa} as well as position space \cite{Wagner22,Dabrowski:2020ixn,Petruzziello:2021vyf,Wagner:2021thc}.

In this work, we find the dual description to the linear and quadratic GUP (LQGUP) \cite{Ali:2009zq}, generalizing recent work on the purely quadratic GUP \cite{Wagner21a}. The ensuing curvature tensor in momentum space is proportional to the non-commutativity of the coordinates in the original description.

\section{Momentum space curvature from LQGUP}

The LQGUP is characterized by a commutation relation of the form
\begin{equation}
	[\hx^a,\hp_b]=i\delta^a_b\left[1-\alpha l_p\hp+\beta l_p^2\hp^2\right]+i\left[-\alpha l_p\hp+\beta'l_p^2\hp^2\right]\frac{\hp^a\hp_b}{\hp^2},
\end{equation}
with the dimensionless parameters $\alpha,$ $\beta$ and $\beta',$ the Planck length $l_p$ (we choose units in which $\hbar=1$), and where $\hp^2=\delta^{ab}\hp_a\hp_b.$ The Jacobi identities constrain the commutator of the coordinates to read
\begin{equation}
	[\hx^a,\hx^b]\simeq\theta\hat{J}^{ba}\equiv \left(2\beta-\beta'+\alpha\alpha'\right)l_p^2\hat{J}^{ba}\label{NonComPert},
\end{equation}
with the orbital angular momentum $J^{ab}=2\hx^{[a}\hp^{b]},.$

By virtue of Darboux's theorem, it is always possible to describe a non-canonical system like the one at hand in terms of coordinates $\hX^i,$ $\hP_i,$ satisfying the ordinary Heisenberg algebra. Following the approach introduced in \cite{Wagner21a}, assume this transformation to be of the form
\begin{equation}
	\hX^i=e^i_a(\hp)\hx^a,\hspace{1cm}\hP_i=(e^{-1})^a_i(\hp)\hp,\label{trans}
\end{equation}
where the matrix $e^i_a$ reads to second order
\begin{equation}
	e^i_a\simeq\left(1-\alpha l_p\hp+\frac{\beta-\beta'-2\alpha^2}{2}l_p^2\hp^2\right)\delta^i_a+\left[-\alpha \frac{l_p}{\hp}+\left(\beta +\alpha^2\right)l_p^2\right]\hp^i\hp_a.
\end{equation}
As a result, the distance from the origin and the free-particle Hamiltonian become
\begin{align}
	\sigma_x(\hX)^2=&\hx^2=\delta_{ab}(e^{-1})\hX^i (e^{-1})\hX^j,\\
		\hat{H}_{\text{kin}}=&\frac{\hp^2}{2m}=\frac{\hP_ie^i_a\delta^{ab}e_b^j\hP_j}{2m}\equiv\frac{\hP_i\hP_jg^{ij}}{2m}.
\end{align}
Following the logic in \cite{Wagner21a}, these quantities may still be interpreted as distance from the origin and canonical Hamiltonian\footnote{Alternatively, the Hamiltonian may be understood as the geodesic distance from the origin in momentum space requiring a more general transformation than \eqref{trans}, see \cite{Wagner22}. }, albeit now defined on a curved momentum space as described by the \emph{vielbeine} $e^i_a$ and the metric
\begin{equation}
	g^{ij}=\delta^{ab}e_{a}^ie_b^j.\label{GenMomMet}
\end{equation}
Assuming the momentum-space geometry to be described by the Levi-Civita connection (for some subtleties surrounding this choice see \cite{Wagner21a}), we can determine the scalar curvature tensor in momentum space $S^{ikjl}$ 
\begin{equation}
	S^{ikjl}\simeq  2\left(2\beta-\beta'+\alpha^2\right)l_p^2\left(g^{ij}g^{kl}-g^{ik}g^{jl}\right)\propto \theta l_p^2,
\end{equation}
implying that the background is maximally symmetric  as expected from second-order corrections to a flat background. Furthermore, the curvature tensor is directly proportional to the non-commutativity of the coordinates as had already been obtained in \cite{Wagner21a} and \cite{Wagner22}. 

Summarizing, the correspondence between the GUP and curved momentum space as well as the proportionality of momentum-space curvature and non-commutativity of coordinates appear to be robust results.
	
\section*{Acknowledgments}
The author was supported by the Polish National Research and Development Center (NCBR) project ''UNIWERSYTET 2.0. --  STREFA KARIERY'', POWR.03.05.00-00-Z064/17-00 (2018-2022) and  acknowledges networking support by the COST Action CA18108. He is indebted to A. Farag Ali for insightful discussions.


\begin{thebibliography}{xx}
		
		\bibitem{Loll:2022ibq}
		R.~Loll, G.~Fabiano, D.~Frattulillo and F.~Wagner,
		[arXiv:2206.06762 [hep-th]].
		
		%
		\bibitem{Addazi:2021xuf}
		A.~Addazi, \textit{et al.}
		Prog. Part. Nucl. Phys. \textbf{125} (2022), 103948
		[arXiv:2111.05659 [hep-ph]].
		
		\bibitem{Mead:1964zz}
		C.~A.~Mead,
		Phys. Rev. \textbf{135} (1964), B849-B862
		
		\bibitem{Amelino-Camelia:2000stu}
		G.~Amelino-Camelia,
		Int. J. Mod. Phys. D \textbf{11} (2002), 35-60
		[arXiv:gr-qc/0012051 [gr-qc]].
		
		\bibitem{Maggiore:1993rv}
		M.~Maggiore,
		Phys. Lett. B \textbf{304} (1993), 65-69
		[arXiv:hep-th/9301067 [hep-th]].
		
		\bibitem{Kempf:1994su}
		A.~Kempf, G.~Mangano and R.~B.~Mann,
		Phys. Rev. D \textbf{52} (1995), 1108-1118
		[arXiv:hep-th/9412167 [hep-th]].
		
		\bibitem{Bosso:2020aqm}
		P.~Bosso,
		Class. Quant. Grav. \textbf{38} (2021) no.7, 075021
		[arXiv:2005.12258 [gr-qc]].
		
		\bibitem{Bosso:2022vlz}
		P.~Bosso, L.~Petruzziello and F.~Wagner,
		[arXiv:2206.05064 [gr-qc]].
		
		\bibitem{Buoninfante:2019fwr}
		L.~Buoninfante, G.~G.~Luciano and L.~Petruzziello,
		Eur. Phys. J. C \textbf{79} (2019) no.8, 663
		[arXiv:1903.01382 [gr-qc]].
		
		\bibitem{Kowalski-Glikman:2003qjp}
		J.~Kowalski-Glikman and S.~Nowak,
		Class. Quant. Grav. \textbf{20} (2003), 4799-4816
		[arXiv:hep-th/0304101 [hep-th]].
		
		\bibitem{Wagner21a}
		F.~Wagner,
		Phys. Rev. D \textbf{104} (2021) no.12, 126010
		[arXiv:2110.11067 [gr-qc]].

		\bibitem{Wagner22}
		F.~Wagner,
		[arXiv:2206.04601 [gr-qc]].
		
		\bibitem{Singh:2021iqa}
		R.~Singh and D.~Kothawala,
		Phys. Rev. D \textbf{105} (2022) no.10, L101501
		[arXiv:2110.15951 [gr-qc]].
		
		\bibitem{Dabrowski:2020ixn}
		M.~P.~Dabrowski and F.~Wagner,
		Eur. Phys. J. C \textbf{80} (2020) no.7, 676
		[arXiv:2006.02188 [gr-qc]].
		
		\bibitem{Petruzziello:2021vyf}
		L.~Petruzziello and F.~Wagner,
		Phys. Rev. D \textbf{103} (2021) no.10, 104061
		[arXiv:2101.05552 [gr-qc]].
		
		\bibitem{Wagner:2021thc}
		F.~Wagner,
		Phys. Rev. D \textbf{105} (2022) no.2, 025005
		[arXiv:2111.15583 [gr-qc]].
		
		\bibitem{Ali:2009zq}
		A.~F.~Ali, S.~Das and E.~C.~Vagenas,
		Phys. Lett. B \textbf{678} (2009), 497-499
		[arXiv:0906.5396 [hep-th]].
		
\end{thebibliography}
\end{document}